\documentstyle[11pt,epsfig]{article}
\begin{document}
\baselineskip 14.0pt
\def\oneskip{\vskip\baselineskip}
\def\xr#1{\parindent=0.0cm\hangindent=1cm\hangafter=1\indent#1\par}
\def\la{\raise.5ex\hbox{$<$}\kern-.8em\lower 1mm\hbox{$\sim$}}
\def\ma{\raise.5ex\hbox{$>$}\kern-.8em\lower 1mm\hbox{$\sim$}}
\def\ea{\it et al. \rm}
\def\am{$^{\prime}$\ }
\def\as{$^{\prime\prime}$\ }
\def\msol{M$_{\odot}$ }
\def\kms{$\rm km\, s^{-1} $}
\def\cm3{$\rm cm^{-3} $}
\def\Ts{$\rm T_{*} $}
\def\Vs{$\rm V_{s} $}
\def\n0{$\rm n_{0} $}
\def\B0{$\rm B_{0} $}
\def\ne{$\rm n_{e} $}
\def\Te{$\rm T_{e} $}
\def\Tgr{$\rm T_{gr} $}
\def\Tgas{$\rm T_{gas} $}
\def\Ec{$\rm E_{c} $}
\def\Fh{$\rm F_{h} $}
\def\Hb{H$\beta $}
\def\erg{$\rm erg\, cm^{-2}\, s^{-1} $}

\centerline{\Large{\bf A grid of composite models for the simulation of the}}

\centerline{\Large{\bf emission-line spectra from the NLR of active galaxies}}.

\bigskip

\bigskip

\bigskip

\centerline{ $\rm M. \, Contini^1 \,\, and \,\,\, S. M. \, Viegas^2  $}

\bigskip

\bigskip

\bigskip

$^1$ School of Physics and Astronomy, Tel-Aviv University, Ramat-Aviv, Tel-Aviv,
69978, Israel

$^2$ Istituto Astron\^{o}mico e Geof\' \i sico, USP, Av. Miguel Stefano, 
4200,04301-904
S\~{a}o Paulo, Brazil

\bigskip

\bigskip

\bigskip

\bigskip

\bigskip

\bigskip

\bigskip

Running title : Grid of models for the NLR of Seyfert galaxies spectra

\bigskip

\bigskip

\bigskip

\bigskip

\bigskip

\bigskip

subject headings : galaxies : nuclei - galaxies : Seyfert - shock waves -
model calculation

\newpage

\section*{Abstract}

A grid of composite models for the narrow line region of
active galaxies, which  consistently account for both the effect of a 
photoionizing radiation  from the active center and of a  shock front,
is presented. Theoretical results, calculated with the SUMA code, 
are given for different values of shock velocities,
preshock densities, geometrical thickness of the clouds, and, particularly, for
the ionizing radiation intensity in a large range.
The input parameters are chosen within the ranges indicated by 
previous fits of several observed emission-line and continuum
spectra from active galaxies. 
Shock velocities from 100 \kms ~to 1500 \kms ~and preshock densities
from 100 \cm3 to 1000 \cm3 are considered.
The line intensities of the most 
important ultraviolet, optical and infrared
transitions are obtained and  are listed in several tables.

\newpage

\section{Introduction}

The narrow line region (NLR) of active galactic nuclei (AGN) has a complex structure. 
The emission-line ratios also
indicate that a variety of cloud densities and sizes are present in the
NLR.
The observed full width at half maximum (FWHM) of the emission-line
profiles show that clouds with  relatively high velocity coexist with
those with  lower velocities (by a factor $\geq 10$).  
On the other hand, the radiation intensity, coming from the active center 
and reaching the NLR clouds,
decreases with the distance from the central  source.
Therefore, only in very seldom cases the spectrum emitted from the NLR
can be simulated by using an unique set of parameters. Generally, the emitted
spectrum can be  fitted by a weighted average of
single-cloud spectra showing different
characteristics (Contini 1997, Contini, Prieto \& Viegas 1998a,b,
Contini \& Viegas 1999, Contini \& Viegas 2000,
 Contini, Viegas \& Prieto 2000).

Nevertheless, a grid of model calculations 
can  be helpful to set the
first basic conditions in the different observed regions. Although 
photoionization is the main process powering the NLR,
shocks cannot be neglected in the study of the AGN spectra 
(e.g. Viegas-Aldrovandi \& Contini 1989).
Therefore, the grid of models presented in this paper
 should be used to complete the informations
which were obtained until now from the
diagrams by Osterbrock, Tran \& Veilleux (1992). 
In the present calculations
 the effect of the shock is coupled to the effects of photoionization. 
Usually, in the observed spectra, the line characteristics coming from
the shocked and from the
photoionized zones may not be easily recognizable, since
the two regions may interact through the diffuse radiation. Thus,
the resulting line intensities are not just a sum of the line
intensities produced in each zone, but  correspond to values
that are obtained by self-consistent calculations which account
for the interaction of the two regions when computing the
cloud physical conditions.

These new theoretical data 
enlarge the range of the results given by Viegas-Aldrovandi \& Contini (1989),
which presented a smaller number of emission-lines. In this paper,
the input parameters are chosen within the range inferred
from the fit of different galaxy spectra
(e.g. Contini, Prieto, \& Viegas 1998a,b, Contini \& Viegas 2000,
Contini, Viegas, \& Prieto 2000).

A larger number of emission-lines in the UV-optical-IR
 spectral ranges  are presented because: (1)
different lines from the same multiplet reveal the temperature and
density conditions of the emitting gas; (2) lines produced by 
different ions of a same
element are important to constrain 
 the intensity of the active center (AC) radiation, and (3)
lines from a same ion but in the different wavelength ranges give 
further information on the physical conditions of the emitting 
region.

 The single-cloud simulations  are obtained with
the SUMA code (see, for instance, Viegas \& Contini 1994). Notice that
the simulations apply whether the shocks originate from an interaction
of the emitting clouds with a  radio jet or from a radial
outflow of the clouds across a lower density gas.

\section{The grid}

In our model, the clouds are moving outwards from the galaxy center.
Therefore the shock forms on the outer edge of the cloud whereas the
ionizing radiation reaches the opposite edge, which faces the active center.

A plane-parallel geometry is adopted. In order to calculate the
physical conditions of the emitting  gas, the cloud is
divided in a large  number of slabs. 
In each slab, the fractional abundances of the
ions of each chemical element are calculated by solving the ionization equilibrium
equations, coupled to the energy balance equation. The ionization rates due
to all the ionizing mechanisms, as well as all recombinations rates, are included
in the calculations (see, for instance Osterbrock 1989). In particular, 
the ionization due to the primary radiation (from the central source), 
to the diffuse  radiation, generated by the free-free and free-bound transitions
of the shocked and photoionizad gas, as well as the collisional ionization,  are
all accounted for. 
Because the shock front and the ionizing radiation
act on the opposite edges of the cloud, the calculations imply  
some iterations.
The gas entering the shock front is thermalized  to high temperature,
which depends on the shock velocity. Then, throughout the cloud, in each slab
the temperature is calculated either by the enthalpy
equation in the region close to the shock or by thermal equilibrium
further on, where radiation processes prevail. The cooling and heating rates,
which depend strongly on the gas density, determine the 
temperature in each slab.
The geometrical thickness of the slabs is automatically calculated in order to obtain
a smooth gradient of the temperature throughout the cloud.
The density within each slab is calculated by the compression
equation. Thus, in order to define
the density distribution across the cloud, the first run
starts at the shocked edge. 
The next iteration starts at the inner edge of the cloud
reached by the ionizing radiation from the active center,
ends at the shocked zone, and the physical conditions of the gas
are calculated across the cloud. 
As in  the first iteration, the third iteration starts at 
the shock front. After recalculating the  physical condition across
the cloud, the emission-line and continuum spectra are computed.
Since the density is low, 
we assume that gas is optically thin to the emission-lines.

Depending on the geometrical thickness of the cloud, the diffuse 
radiation may bridge the radiation-dominated and the shock-dominanted
sides. Thus, usually the cloud  models are  
matter-bounded, unless the geometrical thickness is very large, allowing 
for a neutral hydrogen column density larger than $10^{18}$ cm$^{-2}$ both 
in  the shocked and  photoionized zones.

The input parameters for a single-cloud model 
are the shock velocity, \Vs, the preshock density,
\n0, the preshock magnetic field, \B0, the ionizing radiation spectrum,
the chemical abundances,
the dust-to-gas ratio by number, d/g,
and the geometrical thickness of the clouds, D.  A power-law,
characterized by the power index $\alpha$ and the
flux, \Fh, at the Lyman limit, reaching the cloud (in units of cm$^{-2}$
s$^{-1}$ eV$^{-1}$) is generally adopted.
For all the models,  \B0 = $10^{-4}$ gauss, $\alpha_{UV}$ = 1.5,
and $\alpha_X$ = 0.4, 
and cosmic abundances (Allen 1973) are adopted.  Actually, the basic models are 
calculated with d/g = $10^{-15}$.

The emission-line intensities, relative to \Hb,  
are presented in Tables 1 to 12.
Each table corresponds
to a different set of input parameters \Vs, \n0, and D. The results  correspond
to models with 
 increasing ionizing flux,  from the shock-dominated case (SD),
which is  calculated assuming \Fh=0, up to  highly radiation-dominated models
(RD) for which  log \Fh  ~can reach up to 12.48.
The UV and optical emission-lines, relative to \Hb, ~are listed
in the top part of the tables, while the near-infrared (NIR) and IR emission-line ratios
appear in the  bottom. The absolute \Hb ~value is also listed.

Two cases of cloud geometrical thickness are presented: (a)
D=$10^{17}$ cm, which corresponds to   matter-bounded models,
and (b) D=$10^{19}$ cm, corresponding to 
radiation-bounded models. Notice, however, that  for models
with \Vs=500 \kms ~and \n0=300 \cm3, the results for matter-bounded models
are obtained with  
$10^{18}$ cm, because for D=$10^{17}$ cm the cloud  is fully ionized
with T$>$ 3. 10$^6$ K,   reducing the intensity of most of the emission-lines to zero.

Also for models with \Vs=200 \kms and \n0=200 \cm3, the lowest D is
$10^{18}$ cm, in order to find out the critical D in the low velocity-density
models.
In fact, Tables 1 to 4 refer to low preshock densities and  velocities (\n0 = 100 - 200 \cm3,
\Vs = 100 - 200 \kms). Tables 5 to 10 show the results obtained  for different \Vs ~and
a constant \n0 = 300 \cm3. Notice that, besides the adiabatic jump at
the shockfront, the gas is compressed downstream depending on \Vs, 
reaching  the densities which are generally observed in the NLR of AGN.

The results for high velocity shock-dominated clouds appear
in Tables 11 and 12. For these models,
the line ratios are given for different distances from the shock
front, in the region where the gas temperature  is rapidly
decreasing, as shown by the  temperature of the
corresponding slab, which is
also listed in the tables. Notice, however, that the line intensities listed 
correspond  to the result of  the integration across the cloud, 
and not only to the slab contribution.
Except for model 88 (Table 11), which represents
 the case of a high \Vs ~(1500 \kms) and low \n0 (300 \cm3),
all the other models correspond to high \Vs ~and high \n0,
since we are assuming that \Vs ~and \n0 dependence with the distance from
the center follows the observed general trend  from the BLR (high velocity and
density clouds closer to the center) to the NLR (lower velocity and density
farther from the center).           

In Table 13 the maximum value of the 
downstream densities and temperatures are given for
each model, which is identified by the shock velocity (first
column) and by the preshock density (second column). 

In Table 14 some significative line ratios for SD models are presented.
When the  values for  small ($10^{17}-10^{18}$ cm) and large ($10^{19}$ cm) D
are very similar only one value is given. In the opposite case, the first value
refers to small D and the second to large D.

\section{Comments}

The results from SD and RD models are very different.
Particularly, the \Hb ~absolute value is very low in SD models 
compared to RD models.
Recall that photoionization is generally an efficient mechanism
for  ionizing the gas,
while shocks are more  efficient heating the gas.
Thus, the shocked zone is usually at a higher temperature  and has
a  H$^+$ emitting zone  smaller than the photoionized region.
Both conditions do not favor recombination lines, so,  the 
H lines are less intense, and the luminosity of \Hb ~coming
from the shocked gas is lower.

The RD models are composite,
because they contain also the contribution of  the shocked gas.
The calculated absolute line intensity corresponds to the value at the
NLR. Any average spectrum  must be obtained from 
weighted single-cloud models. The weights correspond to the
single-cloud areas. In order to compare to the observed 
(at Earth) absolute
flux, the dilution due to the distance must be taken into account.

\subsection{SD models}

\centerline{\it UV lines}

For \Vs = 200 - 300 \kms, OVI/\Hb  ~is  maximum . The temperatures 
downstream are 6 $10^5$ - 1.4 $10^6$ K, whereas
the postshock densities ($\geq 10^3-10^4$ \cm3) are  low enough and
the cloud has a large high temperature zone. For the emission-lines 
 NV, SiIV, OIV, NIV] and CIV, and SiIII, relative to \Hb, the maximum
value occurs for \Vs=200 \kms. However,  in the matter-bounded case
(D=$10^{17}$ cm), the line ratios are also 
high for \Vs=100 \kms, since models where a large zone of emitting gas 
at T$\geq 10^4$ K is cut out from the cloud  favor  the
high ionization lines as compared to \Hb.

The models with those velocities  are radiation-bounded  for D slightly larger 
than D=$10^{18}$ cm. In fact, the results  shown in Tables 3 and 4
are very similar. The values of Ly$\alpha$/\Hb 
~are very sensitive to D and have a maximum at \Vs=200 \kms. On the other hand,
HeII 1640/\Hb ~is rather low in SD models and reaches a maximum value 
of 2.7 for \Vs=300 \kms.

\centerline{\it Optical lines}

The lowest value of the  [OIII] 5007+4959/[OII] 3727 line ratio 
is shown by SD models (see Table 14), whereas 
 the lines from the  first ionized ions ([NII], [SII], FeII,
SiII, [SiII], etc.) are generally strong. The [OIII]/[OII] 
line ratio becomes larger than unit for \Vs$\geq$ 300 \kms. 

Values of
[OIII]5007+4959/[OIII]4363 line ratio are very low ($<$ 20), which 
are characteristic of LINERs (see, e.g., Contini 1997). This line ratio decreases 
with increasing temperature  and density of the emitting gas. Table 14
shows an opposite trend, i.e. an increasing line ratio with increasing \Vs ~and \n0. 
This is due to the fact that, at higher \Vs, 
higher ionization states of O, collisionally ionized,
 dominate the postshock zone, which is at higher
temperature, whereas the 
 $\rm O^{+2}$ ion is mainly present  in the gas
at  lower temperature, where ionization by the diffuse radiation dominates.

Neutral lines  such as [OI] and [NI] lines, are relatively strong. On the
other hand,  high-ionization lines, relative to \Hb, as 
[FeVII, [FeX] and [FeXI] peak at  \Vs=300 \kms, \Vs $>$ 500 \kms, 
and \Vs ~= 1500 \kms, respectively. 

Another important issue, particularly raised  in the case
of NGC 4151 (Contini, Viegas, \& Prieto, 2000), refers to the 
[OIII]5007+4959/[OII]3727 line ratio. It is argued whether this ratio
increases faster with \n0 (for a fixed \Vs = 200 \kms) or with \Vs 
~(for a fixed \n0 = 200 \cm3). 
In the latter case the density increases downstream because compression increases
with \Vs. It can be seen in 
Fig. 1 that the slope is flatter for models calculated
as function of the preshock density \n0. This indicates that, in the range of   
the corresponding downstream densities, compression due to \Vs  ~prevails
over the effect of a higher \n0 . As  shown in Table 13, a model with a higher \n0 
(300 \cm3) and a lower \Vs ~(100 \kms) leads to a lower
density downstream than a model with a higher \n0 (200 \cm3) and lower \Vs
~(200 \kms).
Therefore, lower [OIII]/[OII]  corresponds to a  lower \n0 
and/or to a lower \Vs.

\centerline{\it IR lines}

Most of the coronal lines are  in the NIR range.  
They become observable for \Vs $\geq$200 \kms ~and 
their intensities increase with \Vs
~reaching  a maximum at 1500 \kms. On the other hand,
lower ionization lines are strong at low \Vs. Notice that 
high values for  [FeII]26/\Hb, [SiII]34.8/\Hb, 
[NeII]12.8/\Hb, and [OI]63/\Hb ~are obtained with 
\Vs=100 \kms, \n0=100 \cm3, and D=$10^{19}$ cm (Table 2).
These input parameters favor the formation of a large 
low-ionization zone in a shocked cloud.
We have chosen the infrared [SiIX]/[SiVII] and [NeIII]/[NeII] line ratios 
 as indicative of SD models (see Table 14).

\centerline{\it High velocity models}

High velocity shocks correspond to high temperatures in the
postshock region. For 1000 \kms ~and 1500 \kms, the temperature
reaches 1.5 $10^7$ K and 3.5 $10^7$ K, respectively. If present
in the NLR, such high velocity clouds may
contribute to the observed X-ray emission-lines and  soft X-ray
continuum. By cutting the cloud at different distances from the shock front,
different matter-bounded models can be obtained and
very different spectra can be generated, because the fractional abundances of
the ions of a given element peak  at different temperatures.
 
However, the
strong compression, due to high \Vs, coupled to high \n0,
leads to a high cooling rate and a drastic change in the temperature.
The jump of the physical conditions
can reduce the volume of gas with temperatures between $10^4$ and $10^6$ K
(Tables 11-12). Thus,
most of the intermediate-ionization lines can be drastically reduced.
Recall that the intensity of a line depends on the volume of gas with the
ideal physical conditions to emit it. 

\subsection{RD composite models}

Lines from different ionization stages peak for different values of 
\Fh. However, in some cases, the line intensity ratios have no simple 
dependence on \Fh ~because the effect of the shock, modifying the 
stratification of the ions downstream (see, e.g., Contini \& Viegas 1991),
interacts with the effect of the primary and diffuse ionizing radiation.

High ionization lines (e.g., OVI and NV in the UV, coronal lines
in the NIR, etc.) can be strong only for very high \Fh ~and \Vs  $\geq$ 200 \kms
~(e.g., Table 4, model 39, and Table 9, model 76).

In the optical range, [FeVII]/\Hb ~peaks at the highest value 
of \Fh ~except for the models with \Vs=100 \kms
~where the maximum is reached for log(\Fh) = 10.48 and 11.48 whether
 D is equal to $10^{17}$ cm or to $10^{19}$ cm, respectively.  
Notice that [FeX] and [FeXI] follow a similar trend. 
As expected, the infrared low ionization [FeII] 26 line  
decreases with \Fh ~for all \Vs.

The most significant optical line ratios 
are given as function of log \Fh ~for different values of \Vs
~in Figs. 2, 3, and 4.  The
[OIII]/[OII] ratio is basically used for modeling, as an 
ionization indicator, because
both lines are generally observed and strong.
Notice the increasing trend of this line ratio
 with increasing \Fh, for all models
with \Vs $<$ 300 \kms. The same trend appears for models with \Vs$\geq 300$ \kms
and high D. 
Comparing to the results plotted in Fig. 1, which refers to SD models,
we see that  the [OIII]/[OII] line ratio
is much more sensitive to the intensity of the flux radiation from the AC
than to the shock velocities.

The [SII] line ratio  reveals the density of the emitting gas.
Because of the low  critical density of the [S II] lines,
only models with \n0=100 and 200 \cm3 can have [SII]6717/6730 $\geq$ 1
(Fig. 3).

The [OIII]5007+4959/[OIII]4363 line ratio is a temperature indicator 
and is used to distinguish between shock dominated and radiation
dominated models (Fig. 4). 

When the effect of photoionization and shock are both  taken
into account, some results may not be so obvious. For example,
the behaviour of the [O III] line ratio, which can be  higher 
for higher \n0 ~and \Vs, as shown in Fig. 4, or  the
behavior of the [Fe VII] 6087 with D.

For sake of clarity and understanding, the distributions of 
 the electron density and temperature across the 
 cloud are shown in Figs. 5a, 6a, and 7a for three different models
(top panel), as well as the ion fractional abundances of
 H$^{+1}$,  O$^{+2}$, Fe$^{+6}$, Fe$^{+9}$, and Fe$^{+10}$
 (bottom panel).
Some ionization rates are plotted in Figs. 5b, 6b, and 7b.

The shockfront is on the right and the cloud edge reached by the photoionizing
flux on the left.

Figs. 5a and b refer to model 69 (\Vs=300 \kms, \n0=300 \cm3, D=$10^{19}$ cm, 
and log \Fh=12).  These figures can help to understand 
the rather high [OIII] 5007+4959/4363 line ratios found in some models.
It can be seen that 
the [OIII] lines are emitted from a region where a lukewarm gas 
is mainly ionized  by  primary and diffuse radiation, while
higher ionization ions, like  Fe$^{+6}$, Fe$^{+9}$, and Fe$^{+10}$,
appear at the high temperature postshock zone.

Fig. 5c refers to the SD model 63 (\Vs=300 \kms, \n0=300 \cm3).
 The [OIII] lines are emitted in
the region where  the high postshock temperature is rapidly dropping  to lower values.
In this case  5007/4363 ratio is very low ($\sim$ 12)
reflecting the shock effect on the gas temperature. Comparison to
Fig. 5a,  where the bulk of O$^{+2}$ is at lower temperature, 
can explain the behavior of the 5007/4363 ratio in RD and SD models.

Figs. 6 and 7 correspond to models 28 (D=$10^{18}$ cm) 
and 37 (D=$10^{19}$ cm), 
respectively, both  with  \Vs=200 \kms, \n0=200 \cm3, and log \Fh=11.78,
illustrating the differences for narrow and wide cloud models.
Notice that for  both models
 an extended zone of  lurkewarm gas  (T of the order of 10$^4$ K)
appears in the internal region of the cloud. However, on closer inspection,
comparing the temperature and O$^{+2}$ distributions in Figs 6 and 7, 
we see that the [OIII] lines are
emitted from a gas at a lower tamperature 
in a wide cloud than in a narrower cloud. Therefore,
the [OIII] 5007/4636 line ratio is lower for small D models.

When the shock velocity is high enough, the high ionization lines come from 
both sides of the cloud, i.e., from the hot postshock zone and from the
slabs closer to the photoionized edge. Because the high ionization zones 
are closer to the edges, their volumes are less affected than the volume of
the low ionization zone 
by a decrease of the cloud geometrical thickness. We can say that
the emitting volume of the low ionization lines decreases faster
with the geometrical thickness of the cloud
than that of the high ionization lines. On the other hand, \Hb ~is produced
across the whole cloud, thus the high-ionization line
intensity, relative to \Hb ~increases while the \Hb ~absolute flux decreases.
As an example, we can compare the results of [Fe VII] 6087 given
by models 28 and 37, which differ only by the value of D, respectively,
10$^{18}$ cm and 10$^{19}$ cm. The [Fe VII]/\Hb ~ratio decreases from model
28 to 37, while \Hb ~absolute flux increases. The net effect is
a decrease of the absolute flux of [Fe VII] from model 28 to 37.

Figs. 6b and 7b show that in the immediate postshock region collisional
ionization prevails because of  its exponential dependency on the temperature.
After the  rapid decrease of the temperature
downstream, the primary ionization rate dominates throughout the cloud.
The fractional abundance of Fe$^{+6}$ is never  negligible throughout 
the whole cloud
for small D (Fig. 6a), whereas it can be   low in the internal
zone of a wide  cloud (model 37, D=10$^{19}$ cm), depending on the
primary radiation rate.
The temperatures reached  in the radiation dominated side (left) of the cloud
are lower in the wide cloud  model ($\leq$ 1.82 $10^4$ K)
than in the narrower one ($\leq$ 2.23 $10^4$ K).  Both values are
 in a  range which  is crucial with respect to 
rapid  variations in the interplay between ionization and recombination coefficients.

\section{Suggestions}

Let us conclude by describing the method we have used in the last
few years in order to successfully modeling  the NLR spectra of Seyfert galaxies.

First, we chose the \Vs ~corresponding to the FWHM of the observed line
profiles. On this basis, we chose the model fitting  the most important
lines, which are usually those with the lowest observed uncertainty. 
Then, we add other models with different weights to improve the
fitting of all the other observed emission-lines.
Following this procedure, we obtain a picture of the physical conditions 
in the NLR of the galaxy.
Notice, however, that each galaxy is a special case. So, this grid must be used
only as a first approach to explain the observed spectrum. 

In order to have a full description of the NLR, the models fitting  the
emission-line spectrum must also explain the observed continuum spectrum.
This may force the inclusion of new components (clouds under different
physical conditions), which must be tested against the observed 
emission-line spectrum. The process may be repeated until both the
emission-line and continuum spectra are fitted. Once the fit
is achieved, all the information about the clouds present in the NLR
is available, including the relative importance of their contribution
to the emission  in each spectral range.

\bigskip

Acknowledgements: We are grateful to the referee for enlightening comments.
This paper is partially supported by CNPq (304077/77-1),
 PRONEX/Finep (41.96.0908.00), and FAPESP (97/13861-4).

\newpage

{\bf References}

\bigskip

\vsize=26 true cm
\hsize=16 true cm
\baselineskip=14 pt
%
\def\ref {\par \noindent \parshape=6 0cm 12.5cm 
0.5cm 12.5cm 0.5cm 12.5cm 0.5cm 12.5cm 0.5cm 12.5cm 0.5cm 12.5cm}

\ref Allen, C.W. 1973, Astrophysical Quantities (London: Athlone)
\ref Contini, M. 1997, A\&A, 323, 71
\ref Contini, M. \& Viegas, S.M. 1991, A\&A 251, 27  
\ref Contini, M. \& Viegas, S.M. 1999 ApJ, 523,114
\ref Contini, M. \& Viegas, S.M. 2000 ApJ, 535,
\ref Contini, M., Prieto, M.A., \& Viegas 1998a ApJ , 492, 511
\ref Contini, M., Prieto, M.A., \& Viegas 1998b ApJ , 505, 621
\ref Contini, M. Viegas, S.M., \& Prieto, M.A. 2000 ApJ submitted 
\ref Nelson, C.H. et al astro-ph/9910019
\ref Osterbrock, D. E.  1989, "Astrophysics of Gaseous Nebulae and
Active Galactic Nuclei" (University Science Books)
\ref Osterbrock, D.E., Tran, K.D., \& Veilleux, S. 1992, ApJ, 389, 1960
\ref Viegas-Aldrovandi, S.M. \& Contini, M. 1989, ApJ 339, 689 
\ref Viegas, S.M. \& Contini, M. 1994, ApJ, 428, 113  

\newpage

{\bf Figure Caption}

\bigskip

Fig. 1 :

Emission line ratios as a function of \Vs ~(dashed lines, labelled by \n0) 
and \n0 (solid lines, labelled by \Vs). Thin lines correspond to 
[OIII]5007+4959/[OII]3727  and  thick lines to [OII]3727/\Hb.

\bigskip

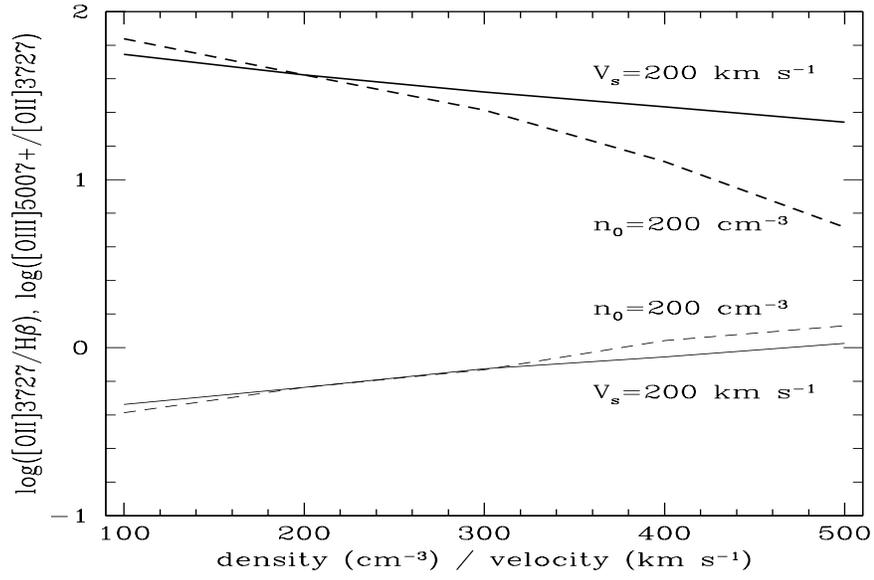
Fig. 2 :

The [OIII]5007+4959/[OII]3727 emission-line ratio versus log \Fh, for
models with:
\Vs=100 \kms and \n0=100 \cm3 (solid lines),
\Vs=100 \kms and \n0=300 \cm3  (dot-dashed lines),
\Vs=200 \kms (dotted lines),
\Vs=300 \kms (short-dashed lines), and 
\Vs=500 \kms (long-dashed lines).
Thin lines refer to results for  D=$10^{17}$-$10^{18}$ cm, 
whereas thick lines, to D=$10^{19}$ cm.

\bigskip

Fig. 3 :

The [SII]6717/[SII]6730 emission-line ratio versus log \Fh.
Same notation as in Fig. 2.

\bigskip

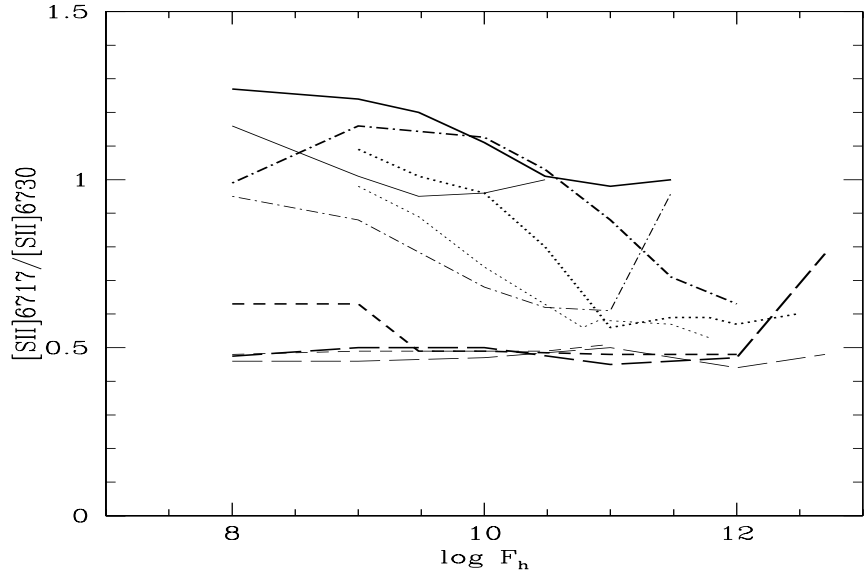
Fig. 4 :

The [OIII]5007+4959/[OIII]4363 emission-line ratio versus log \Fh.
Same notation as in Fig. 2.

\bigskip

Fig. 5

Results from Model 69 (\Vs = 300 \kms, \n0=300 cm$^{-3}$, D=$10^{19}$ cm, 
log \Fh=12). The shocked edge is on the right and the
ionized edge on the left.
(a) : the distribution across the cloud 
of the temperature (dashed line) and density (solid line) in the top panel,
 and of the fractional abundances
of the H$^{+1}$ (solid),  O$^{+2}$ (dot-dashed), Fe$^{+6}$ (dotted),
 Fe$^{+9}$ (short dashed), and Fe$^{+10}$ (long-dashed) 
ions (bottom panel).

(b) :  the distribution across the cloud of  the collisional (dotted), 
diffuse from the shockfront (long-dashed), diffuse from the
edge reached by the photoionizing radiation from the center (dash-dotted), 
and primary (short-dashed)
ionization rates for the O$^{+2}$ ion.

 (c) : same as for (a) in the shock dominated case.

\bigskip

Fig. 6

Results from Model 28 (\Vs=200 \kms, \n0=200 cm$^{-3}$, D=$10^{18}$ cm, 
log \Fh=11.78). (a) Same notation as in Figs. 5a; (b) same notation
as in Fig. 5b, but for Fe$^{+6}$ ionization rates.

\bigskip

Fig. 7

Results from Model 37(\Vs = 200 \kms, \n0=200 cm$^{-3}$, D=$10^{19}$ cm, 
log \Fh=11.78). Same notation as in Figs. 6.

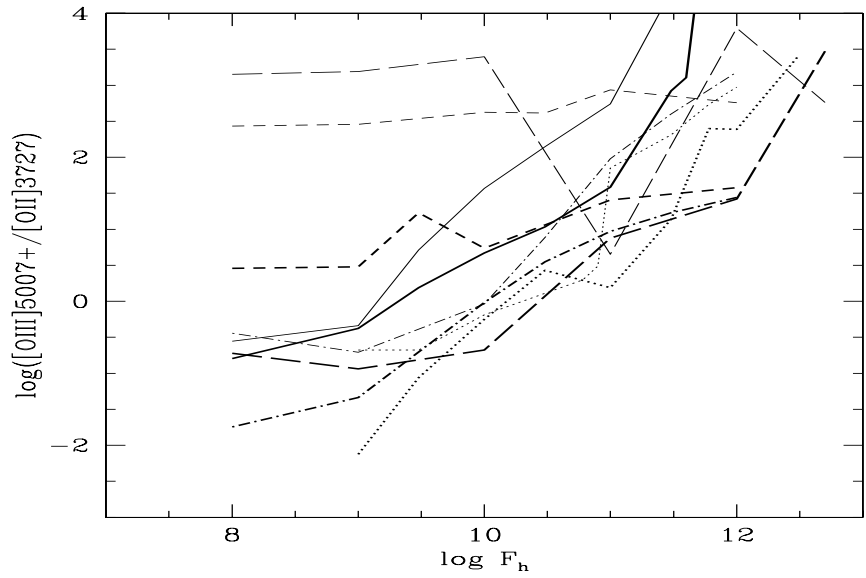
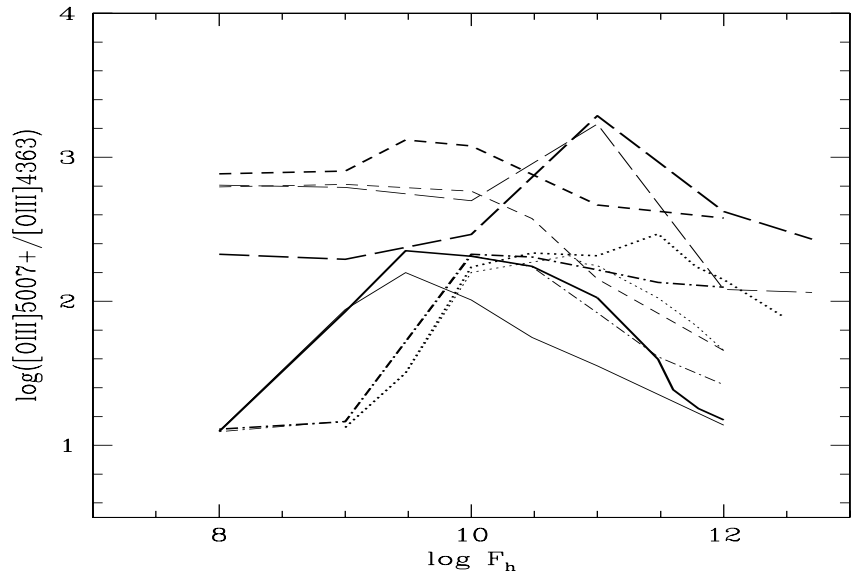
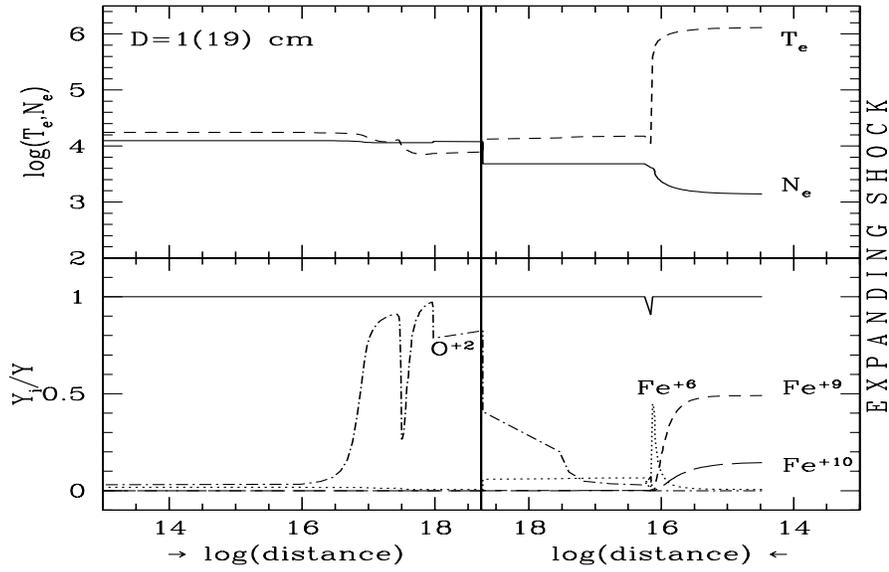
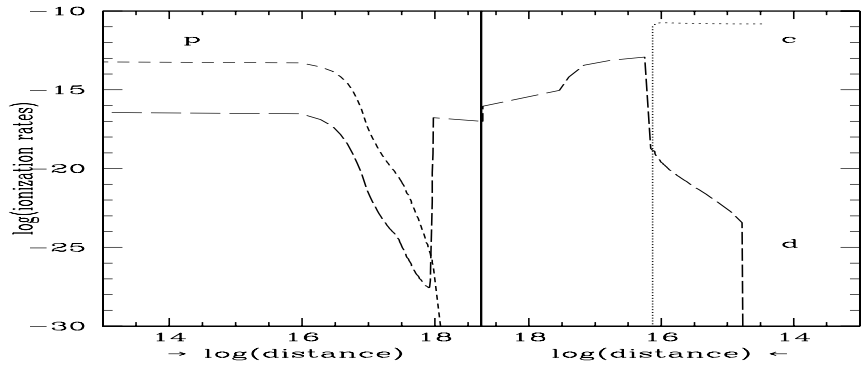
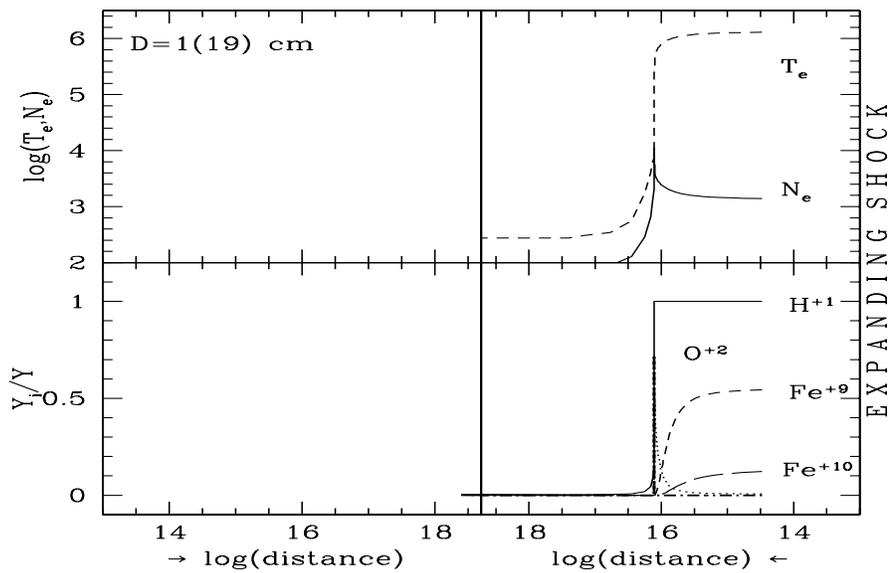
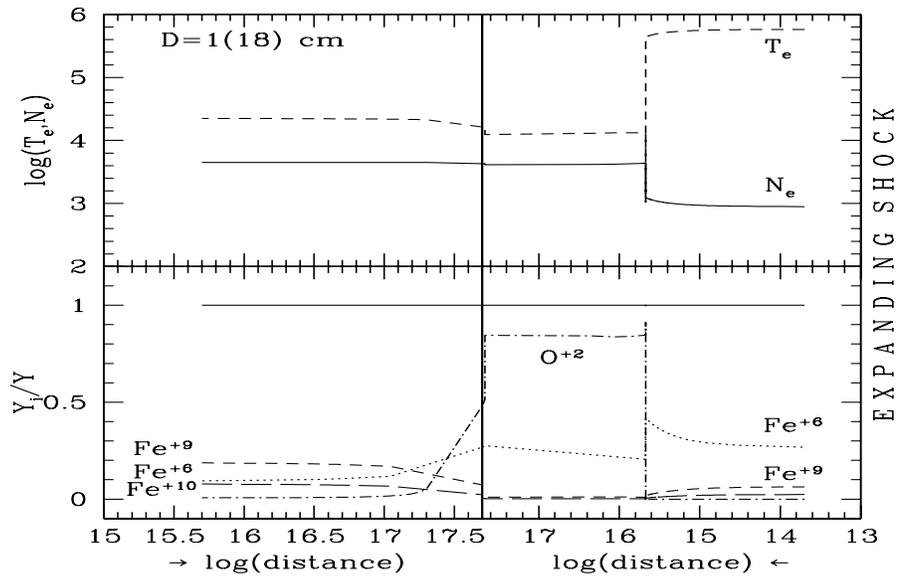
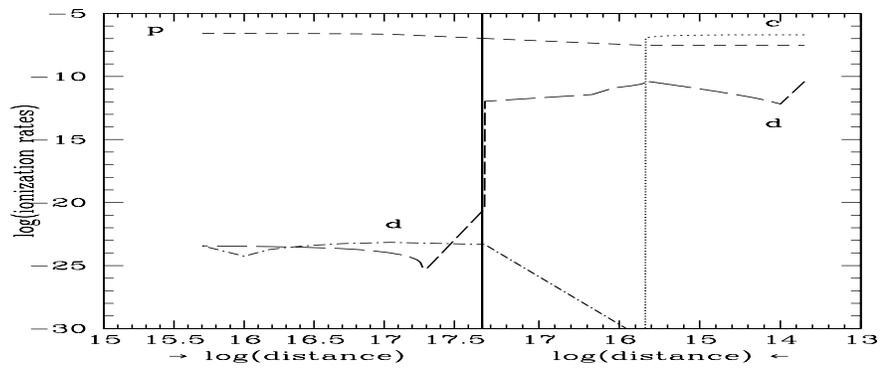
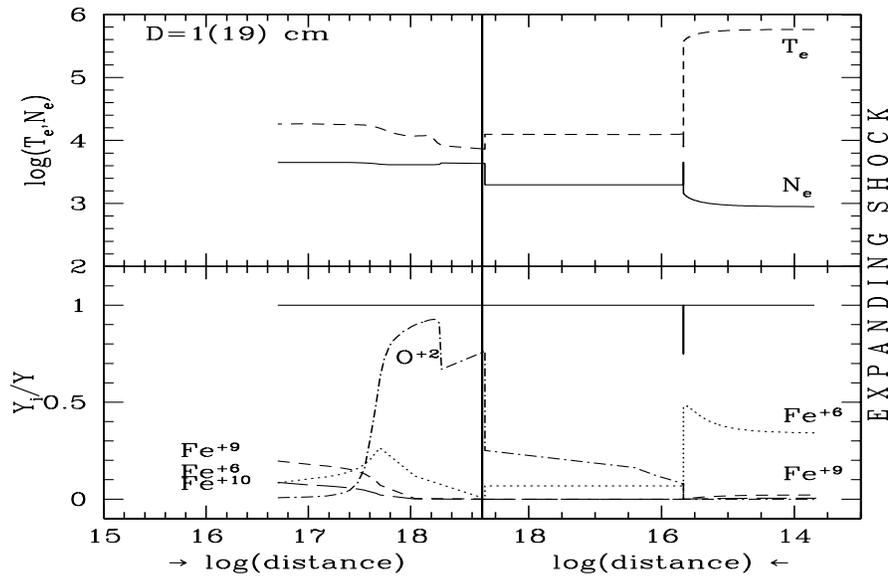
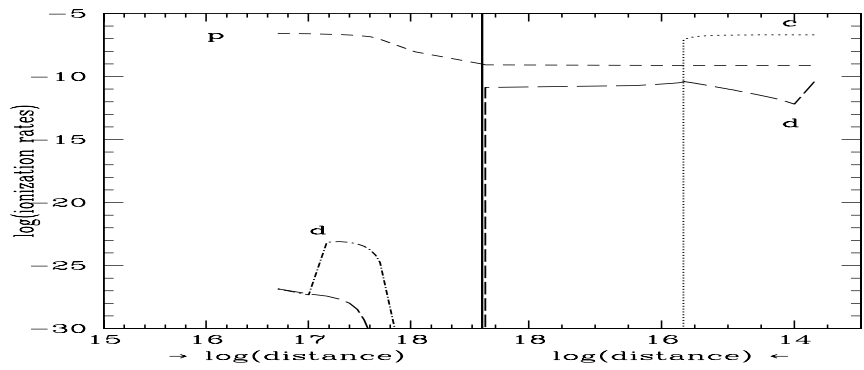
\end{document}